\RequirePackage{ifpdf}
\ifpdf 
\documentclass[pdftex]{sigma}
\else
\documentclass{sigma}
\fi

\begin{document}
\def\omt{\tilde{\omega}}
\def\ti{\tilde}
\def\o{\Omega}
\def\bchi{\bar\chi^i}
\def\In{{\rm Int}}
\def\ba{\bar a}
\def\w{\wedge}
\def\ep{\epsilon}
\def\k{\kappa}
\def\Tr{{\rm Tr}}
\def\ST{{\rm STr}}
\def\ss{\subset}
\def\rn{\vert \alpha\vert^2}
\def\bi{\bibitem}
\def\ot{\oti\def\om{\omega}
\dmes}
\def\bc{{\bf C}}
\def\bz{{\bf Z}}
\def\ptp{\stackrel{\otimes}{,}}
\def\br{{\bf R}}
\def\de{\delta}
 \def\bt{\beta}
 \def\ve{\vert}
\def\al{\alpha}
\def\la{\langle}
\def\ra{\rangle}
\def\Ga{\Gamma}
\def\st{\stackrel{\wedge}{,}}
\def\stv{\stackrel{\wedge}{\vert}}
\def\th{\theta}
\def\lm{\lambda}
\def\U{\Upsilon}
\def\jp{{1\over 2}}
\def\js{{1\over 4}}
\def\d{\partial}
\def\tr{\triangleright}
\def\trl{\triangleleft}
\def\d{\partial}
\def\bq{\}_{P}}
\def\D{{\cal D}}
\def\A{{\cal A}}
\def\G{{\cal G}}
\def\K{{\cal K}}
\def\H{{\cal H}}
\def\P{{\cal P}}
\def\N{{\cal N}}
\def\R{{\cal R}}
\def\B{{\cal B}}
\def\T{{\cal T}}
\def\bT{\bar{\cal T}}
\def\F{{\cal F}}
\def\n{{1\over n}}
\def\si{\sigma}
\def\ta{\tau}
\def\ov{\over}
\def\l{\lambda}
\def\L{\Lambda}
\def\lpb{{\bf \{}}
\def\rpb{{\bf \}}}
\def\ms{\medskip}
\def\noi{\noindent}

\def\pih{\hat{\pi}}

\def\U{\Upsilon}
\def\e{\varepsilon}
\def\bt{\beta}
\def\ga{\gamma}
\def\om{\omega}
\def\be{\begin{equation}}
\def\ee{\end{equation}}
\def\bea{\begin{eqnarray}}
\def\eea{\end{eqnarray}}
\def\D{{\cal D}}
\def\C{{\cal C}}
\def\G{{\cal G}}
\def\H{{\cal H}}
\def\R{{\cal R}}
\def\B{{\cal B}}
\def\K{{\cal K}}
\def\T{{\cal T}}
\def\S{{\cal S}}
\def\bT{\bar{\cal T}}
\def\F{{\cal F}}
\def\n{{1\over n}}
\def\si{\sigma}
\def\ta{\tau}
\def\ot{\otimes}
\def\l{\lambda}
\def\L{\Lambda}
\def\ve{\vert}
\def\nr{\nabla_R}
\def\nl{\nabla_L}
\def\pih{\hat{\pi}}

\def\e{\varepsilon}
\def\bt{\beta}
\def\ga{\gamma}

\renewcommand{\PaperNumber}{003}

\FirstPageHeading

\renewcommand{\thefootnote}{$\star$}

\ShortArticleName{Af\/f\/ine Poisson Groups and WZW Model}

\ArticleName{Af\/f\/ine Poisson Groups and WZW Model\footnote{This
paper is a contribution to the Proceedings of the Seventh
International Conference ``Symmetry in Nonlinear Mathematical
Physics'' (June 24--30, 2007, Kyiv, Ukraine). The full collection
is available at
\href{http://www.emis.de/journals/SIGMA/symmetry2007.html}{http://www.emis.de/journals/SIGMA/symmetry2007.html}}}

\Author{Ctirad KLIM\v C\'IK}
\AuthorNameForHeading{C. Klim\v c\'\i k}

\Address{Institute de math\'ematiques de Luminy,
 163, Avenue de Luminy, 13288 Marseille, France}
\Email{\href{mailto:klimcik@iml.univ-mrs.fr}{klimcik@iml.univ-mrs.fr}}

\ArticleDates{Received October 31, 2007; Published online January 11, 2008}

\Abstract{We give a detailed description of a dynamical system which enjoys a Poisson--Lie
symmetry with two non-isomorphic dual groups. The system is obtained by taking
the $q\to\infty$ limit of the $q$-deformed WZW model and the understanding of its symmetry structure
results in uncovering an interesting duality of its exchange relations.}


\Classification{81T40}

\vspace{-2mm}

\section{Introduction}

Poisson--Lie symmetry of a dynamical system is a generalization
of the standard concept of the Hamiltonian symmetry.  Poisson--Lie symmetric dynamical
system possess a distinguished subalgebra of observables
which is isomorphic to the Poisson algebra of functions on the so-called af\/f\/ine Poisson group.
Given an af\/f\/ine Poisson group, it is not dif\/f\/icult to {\it construct} a~Poisson--Lie symmetric
dynamical system, by using the theory of Heisenberg doubles  or, in other terminology, symplectic
grupoids \cite{Lu}.  However, it is certainly more interesting to {\it discover} a Poisson--Lie
symmetry in a model which was proposed or studied for reasons independent from the
symmetry considerations. This is what exactly
happened when we have studied a $\e\to\infty$ limit of the $\e$-deformed WZW model.
The Poisson--Lie symmetry, which we have discovered in the limit, appears in turn to be a very useful structure for the quantitative analysis of the model. In particular, we shall see that it permits
to identify a remarkable duality of the exchange relations of the theory.

 This paper is a short review of the original research article \cite{K06} and it is written with
the goal to streamline the presentation by omitting various technical details.  Its plan is as follows:
f\/irst we give a compact  introduction  to the concepts of the af\/f\/ine Poisson group and of the Poisson--Lie
symmetry and then we describe the Poisson--Lie symmetry of the $\e$-deformed WZW model for f\/inite $\e$.
In the subsequent Section~4, we identify the Poisson--Lie symmetry
of the $\e\to\infty$ limit of the $\e$-deformed WZW model and, f\/inally, the  Section~5 will be devoted
to the duality of the exchange relations.

\vspace{-1mm}

\section[Poisson-Lie symmetry]{Poisson--Lie symmetry}

 Let $\Pi_B$ denote a Poisson bivector on a Lie group manifold $B$.
As it is well-known \cite{Do,MM,Ko,DS}, $\Pi_B$  gives rise to the Lie algebra
 structure on the space
  $\Omega^1(B)$
consisting of smooth  $1$-form f\/ields~on~$B$:{\samepage
\begin{gather*}
 \{xdy,udv\}_{\Omega(B)}=xud\{y,v\}_B  +x\{y,u\}_Bdv +u\{x,v\}_Bdy,
 \end{gather*}
where  $x$, $y$, $u$, $v$ are smooth functions on $B$ and
$\{y,v\}_B\equiv \Pi_B(dy,dv)$.}

 If, at the same time, the spaces of right and left invariant $1$-form
f\/ields are respectively Lie subalgebras ${\cal G}_L$ and
${\cal G}_R$  of the Lie algebra
$(\Omega^1(B),\{\cdot,\cdot\}_\Omega)$, we say that $B$ is equipped with an
{\it affine Poisson structure} $\Pi_B$~\cite{DS}.

 A {\it dynamical system} is a triple $(P,\Pi_P,H)$, where $\Pi$ is a non-degenerate bivector
 f\/ield on a~manifold $P$ and $H$ is a distinguished function on $P$ called the Hamiltonian.

 {\it Poisson--Lie symmetry} of a dynamical system  $(P,\Pi_P,H)$ is any
smooth map $\mu$ from $(P,\Pi_P)$ into an af\/f\/ine Poisson group
$(B,\Pi_B)$ such that:

  1) the $\mu$-pull-backs of  any $x,y\in {\rm Fun}\,(B)$  satisfy
\[\{\mu^*x,\mu^*y\}_P=\mu^*\{x,y\}_B;\]

 2) for any $x\in {\rm Fun}\,(B)$ it exists $y\in {\rm Fun}\,(B)$ fulf\/illing
 \[\{H,\mu^*x\}_P=\mu^*y.\]
 The map  $\mu$ is often called a {\it moment map} and it induces at the same time the actions of
both  Lie algebras~$\G_L$ and~$\G_R$ on $P$. Indeed, to any $\lm\in\G_L$ it corresponds
 the vector f\/ield
$\Pi_P(\cdot,\mu^*\lm)$  acting on functions on $P$ and to any $\rho\in\G_R$ it  corresponds
$\Pi_P(\cdot,\mu^*\rho)$.  The Lie algebras~$\G_L$ and~$\G_R$ are then interpreted as
the symmetry algebras of the dynamical system.

 A Poisson--Lie symmetry   is called {\it equivariant}  if the af\/f\/ine
 Poisson bivector $\Pi_B$  vanishes at the unit element $e_B$ of the group $B$.
If the symmetry is
equivariant then Lie algebra $\G_L$ is necessarily isomorphic to the Lie algebra $\G_R$.
However, even in this
case the $\G_L$ action on $P$ need not to coincide with the $\G_R$ action.
Non-equivariant symmetry is called {\it anomalous}.
In the presence of the anomaly, $\G_L$ need not be even isomorphic to $\G_R$.

 \section[$\e$-deformed WZW model]{$\boldsymbol{\e}$-deformed WZW model}

 Let $G$ be a connected and simply connected simple compact Lie group, $G^\bc$ its complexif\/ication
 and $LG^\bc$  the group of loops, i.e.\ the group  of smooth maps  from a circle $S^1$ into $G^\bc$.
 Let $B$ be a subgroup of $LG^\bc$  containing only the loops which
are boundary values of holomorphic maps $H(z)$ from the
 southern Riemann hemisphere into the complex group $G^\bc$.
 Moreover, it is required that the value of the $H(z)$ at the south pole
$z=0$ lies in the subgroup
$AN$ of $G^\bc$. (Remind that $AN$ is the subgroup of $G^\bc$ determined by the Iwasawa
decomposition $G^\bc=GAN$.) Set
\[L(\si)\equiv H(e^{i\si})(H(e^{i\si}))^\dagger,\]
where $\dagger$ stands for the standard Hermitian conjugation. In \cite{RS},
Reshetikhin and Semenov-Tian-Shansky  have introduced the following
    af\/f\/ine Poisson
 structure on $B$:
 \begin{gather}
 \{L(\si)\ptp L(\si')\}_B=(L(\si)\ot L(\si'))\e\hat r(\si-\si')-(1\ot L(\si'))
\e\hat r(\si-\si'+2\pi i\ep)(L(\si)\ot 1)\nonumber\\
\qquad{}+\e\hat r(\si-\si')(L(\si)\ot L(\si'))
-(L(\si)\ot 1)\e\hat r(\si-\si'-2\pi i\ep)(1\ot L(\si')).\label{eq1}
\end{gather}
Here $\e$ is the  deformation parameter and{\samepage
\begin{gather*}
\hat r(\si-\si') =  r+
C\,{\rm  cotg}\jp(\si-\si') ,\\
 r=\sum_{\al > 0}{i\vert \al \vert^2\over 2}(E^{-\al}\otimes
 E^{\al}-E^\al\otimes E^{-\al}),\\
 C=\sum_\mu H^\mu\otimes H^\mu+
\sum_{\al >0}{\vert \al \vert^2\over 2}(E^{-\al}\otimes E^{\al}+
E^\al\otimes E^{-\al}).
\end{gather*}
Our conventions for the normalisations of the standard step generators $E^\al$  can be found
in \cite{K06}.}

Reshetikhin and Semenov-Tian-Shansky have interpreted the af\/f\/ine Poisson structure
\eqref{eq1} as the $\e$-deformed current algebra since in the limit $\e\to  0$ it reduces
to the standard Kac--Moody current algebra. It is well-known that the Kac--Moody algebra
underlies the symmetry structure of the so-called chiral WZW model \cite{CGO} which is
a very important dynamical system from the point of view of the conformal
f\/ield theory. Motivated by
this fact,
 Lukyanov and Shatashvili  \cite{LS}
have formulated the following problem: Is there a dynamical system which would be
Poisson--Lie symmetric with respect to the af\/f\/ine Poisson structure \eqref{eq1} and would reduce
to the standard chiral WZW model in the limit $\e\to 0$? The af\/f\/irmative answer to this question
was given in \cite{K04} and it goes along the following lines:

 The space ${\rm Fun}\,(P)$ of the observables of the $\e$-deformed  WZW model  is
${\rm Fun}\,(P)= {\rm Fun}\,(LG)\otimes {\rm Fun}^W(T)$,
where ${\rm Fun}^W(T)$ is the space of Weyl invariant functions on the Cartan torus $T\subset G$.
The non-degenerate Poisson structure on $P$ is determined by the following Poisson brackets
\begin{gather}
\{t\ptp k(\si)\}_P=
 (t\otimes k(\si))(H^\mu\otimes H^\mu), \qquad \{t\ptp t\}_P=0,\nonumber\\
  \{k(\si)\ptp k(\si')\}_P
=(k(\si)\otimes k(\si'))
\hat r_{\ep}(t,\si-\si')
 -\ep\hat r(\si-\si')(k(\si)\otimes k(\si')),\label{eq2}
 \end{gather}
where $ \hat r_{\ep}(t,\si)$ is the well-known Felder  elliptic dynamical $r$-matrix \cite{F}:
\[  \hat r_{\ep}(t,\si)={\ep}\rho\left({\si\over 2\pi},{i\ep}\right)
H^\mu\otimes H^\mu +
{\ep}\sum_{\al}{\rn\over 2}\si_{{-\ep \la
\al,{\rm ln}t\ra}}
\left({\si\over 2\pi},{i\ep}\right)E^\al\otimes E^{-\al}.
\]
The  moment map  $\mu_\ep:P\to B$,
 that realizes the Poisson--Lie symmetry of the $\e$-deformed WZW model,
is then given by a compact formula
\[\mu_\ep(k(\si),t) =Iw_\ep(k(\si)t^{-i\ep}).\]
 Here the  Iwasawa map $Iw_\ep:LG^\bc\to B$  is
 uniquely characterized by the property  that for every $l(\si)\in LG^\bc$ it exists
  $m(\si -i\ep)\in LG$ such that
$l^{-1}Iw_\ep(l)=m(\si)$.

 In \cite{K04}, it was proven the existence of a function $H_\e$ on the phase space $P$
which  gives
the standard chiral WZW Hamiltonian in the limit $\e\to 0$ and fulf\/ils the condition 2) of the def\/inition of the Poisson--Lie symmetry above. We take the function $H_\e$ as
the Hamiltonian of the $\e$-deformed WZW model, knowing that its
 explicit form in variables
$k(\si)$ and $t$ is not available.

\section[The limit $\e\to\infty$]{The limit $\boldsymbol{\e\to\infty}$}

 The $\e\to\infty$ limit of the $\e$-deformed current algebra \eqref{eq1}
reads
\begin{gather}
\{L(\si)\ptp L(\si')\}_\infty=
(L(\si)\ot L(\si'))\hat r(\si-\si')+
  \hat r(\si-\si')(L(\si)\ot L(\si'))\nonumber\\
 \phantom{\{L(\si)\ptp L(\si')\}_\infty=} {} -(L(\si)\ot 1 )(r+iC)
  (1\ot L(\si'))- (1\ot L(\si'))
(r-iC)(L(\si)\ot 1 ).\label{eq3}
\end{gather}
It turns out that  equation \eqref{eq3} def\/ines an af\/f\/ine Poisson structure
on the group $B$, which is  qualitatively dif\/ferent from~\eqref{eq1}.
Indeed, for f\/inite $\e$, the  Lie algebras $\G_L$ and
$\G_R$  associated to the af\/f\/ine Poisson structure~\eqref{eq1}
turn out to be isomorphic
to
each  other and  also to ${\rm Lie}\,(LG)$.
However, in the limit $\e\to\infty$, $\G_L$
is still isomorphic to ${\rm Lie}\,(LG)$ but $\G_R$
is isomorphic to the Lie algebra of another group which we denote as
 $\bar B$. In fact,  $\bar B$ turns out
to be the subgroup of~$LG^\bc$ containing only
 the loops
which
are boundary values of holomorphic maps $\bar H(z)$ from the
northern Riemann hemisphere  into
the complex
group $G^\bc$. Moreover, it is required that the value
of $\bar H$ at the north pole $z=\infty$ lies in the
subgroup $G$ of $G^\bc$.

 The $\e\to\infty$ limit of the exchange relations \eqref{eq2} reads
\begin{gather}
\{e^a\ptp k(\si)\}_\infty= -i(e^a\otimes k(\si))(H^\mu\otimes H^\mu),
\qquad \{a\ptp a\}_\infty=0,\nonumber\\
 \{k(\si)\ptp k(\si')\}_\infty  =(k(\si)\otimes k(\si'))
\hat r_{\infty}(a,\si-\si')
 -\hat r(\si-\si')(k(\si)\otimes k(\si')),\label{eq4}
 \end{gather}
where $a$ parametrizes the Weyl chamber $A$ in ${\rm Lie}\,(T)$ and
\[\hat r_{\infty}(a,\si-\si')
=\sum_\al {i\vert \al \vert^2\over 2}\,{\rm coth}(\al(a))
E^{-\al}\ot E^\al+\C\,{\rm cotg}{\si-\si'\over 2}.\]
\noi There exists a Riemann--Hilbert map $RH:LG^\bc\to B$
that is characterized by the property
$l^{-1}RH(l)\in \overline {B}$ for a  $l\in LG^\bc$.
It turns out that the limiting Poisson brackets \eqref{eq4} do not def\/ine
an everywhere non-degenerate Poisson bivector $\Pi_\infty$ on
 the manifold $LG\times A$.
However, the domain of def\/inition of the map $RH(k(\si)e^a)$
is a dense open submanifold of $LG\times A$ on which
$\Pi_\infty$ is non-degenerate. We denote this domain of def\/inition
 as $P_\infty$.  The Hamiltonian $H_\infty$ is the $\e\to\infty$
limit of
$H_\e$.

 The moment map $\mu_\infty:P_\infty\to B$ that realizes
the Poisson--Lie symmetry of the $\e\to\infty$
limit of the $\e$-deformed WZW model
$(P_\infty,\Pi_\infty, H_\infty)$ is then given by
 \[\mu_\infty(k(\si),a)= RH(k(\si)e^a).\]
It can be shown that  $H_\infty$ fulf\/ils
 the condition 2) of the def\/inition of the Poisson--Lie
 symmetry in Section~2.  Actually, the time
evolution determined by the Hamiltonian
$H_\infty$ is given by a very simple
formula
\[k(\si)\to k(\si-\tau),\qquad a\to a.\]

\section{Duality}

 From the general theory of  Poisson--Lie symmetry presented in Section~2,  we know
that both symmetry  Lie algebras $\G_L={\rm Lie}\,(LG)$ and $\G_R={\rm Lie}\,(\bar B)$ act
on the phase space  $P_\infty$ of the $\e\to\infty$ limit of the $\e$-deformed WZW model.
 Indeed, to any $\lm\in\G_L$  it  corresponds
 the vector f\/ield
$\Pi_\infty(\cdot,\mu_\infty^*\lm)$    and to any $\rho\in\G_R$ it corresponds
$\Pi_\infty(\cdot,\mu_\infty^*\rho)$.  A calculation shows that  the~$\G_L$~action is just
the inf\/initesimal version of
 the action
of $LG$ on itself:
\begin{gather}
 h(\si)\triangleright k(\si)=h(\si)k(\si),\qquad h(\si),k(\si)\in LG,\label{eq5}
 \end{gather}
 but
 the formula for the action of $\G_R$ turns out to be  cumbersome and not
very illuminating. To make this formula more friendly and transparent, we
  shall  parametrize the phase space $P_\infty$ in terms of the group elements $\ti k(\si)\in \bar B$
in such a way the $\G_R$ action
will  be just the inf\/initesimal version of
 the action
of $\bar B$ on itself:
\begin{gather} \ti h(\si)\triangleright \ti k(\si)=\ti h(\si)\ti k(\si),\qquad \ti
h(\si),\ti k(\si)\in \bar B.\label{eq6}
\end{gather}
It turns out that the following simple ``coordinate'' transformation on $P_\infty$ does the job:
\begin{gather}
 \ti k^{-1}(\si)=
e^{-a}k(\si)^{-1}RH(k(\si)e^a),\qquad\ti a=-a.\label{eq7}
\end{gather}
This  relation between the old description of the phase space $P_\infty$ in terms
of  $(a,k(\si)\in LG)$ and the new description in terms of
$(\ti a,\ti k(\si)\in \overline {B})$  entails more  advantages than just the elegant descriptions \eqref{eq5} and \eqref{eq6}
of the both $\G_L$ and $\G_R$ symmetry actions on the phase space.  In fact, it permits
also to characterize $P_\infty$ not just as the domain of def\/inition of the Riemann--Hilbert
map but directly as
$P_\infty= \bar B\times A,$
which means that the inf\/initesimal ${\rm Lie}\,(\bar B)$
action can be lifted to the  {\it global} $\bar B$ action~\eqref{eq6}.

The fundamental exchange relations, that characterize the Poisson algebra
${\rm Fun}\,(P_\infty)$, can be written either in terms of the variables
$(a,k(\si)\in LG)$ or in terms of the dual variables
$(\ti a,\ti k(\si))\in \overline {B})$:
\begin{gather}
\{k(\si)\ptp k(\si')\}_\infty
=(k(\si)\otimes k(\si'))
\hat r_{\infty}(a,\si-\si')
 -\hat r(\si-\si')(k(\si)\otimes k(\si')),\nonumber\\
 -\{\ti K(\si)\ptp \ti K(\si')\}_\infty =(\ti K(\si)\otimes \ti K(\si'))
\hat r_{\infty}(-\ti a,\si-\si')
 -\hat r(\si-\si')(\ti K(\si)\otimes \ti K(\si')).\label{eq8}
 \end{gather}
Here
$\ti K(\si)\equiv \ti k(\si)e^{\ti a}.$
The resemblance of  the  formulae \eqref{eq4} and \eqref{eq8} is remarkable, since it is just the
existence of the dual description of $P_\infty$ which  follows
from the general theory of Poisson--Lie symmetry, but not the
  invariance of the form of the exchange relations under the duality
transformation \eqref{eq7}.   We interpret this form invariance as the $\G_L\leftrightarrow\G_R$ duality
of the chiral $\e\to\infty$ WZW model.

\section{Conclusions}

 The  $\e\to\infty$ limit of the  $\e$-deformed WZW model  is a  dynamical
 system  in which we have discovered the  Poisson--Lie symmetry  with
  two {\it non-isomorphic} symmetry groups $G_L=LG$ and $G_R=\bar B$.
  In particular, the appearance of the group $\bar B$ in the story  is interesting
  because    for  the f\/inite $\e$  WZW model  (including the
standard case $\e=0$) it holds $G_L=G_R=LG$. Thus, in a sense,
 we may  say
 that the symmetry pattern of  the $\e\to \infty$   WZW model
 is richer and more
 intriguing than that of its f\/inite $\e$ counterpart.  The presence of the new symmetry
 group~$\bar B$ turned out to be very useful for the description  of the phase
 space of the limiting model but also for uncovering the interesting duality of the
 exchange relations~\eqref{eq4} and~\eqref{eq8}.  We believe that this duality is one of two
 circumstances which should facilitate
 the program of quantizing the model. The other one is the remarkable fact,
 that in the $\e\to\infty$
 limit the elliptic exchange relations~\eqref{eq2} simplify to the  trigonometric relations~\eqref{eq4}.

\pdfbookmark[1]{References}{ref}
\LastPageEnding

\end{document}